# Cooperation and Underlay Mode Selection in Cognitive Radio Network

Ramy Amer [1], Amr A. El-Sherif [2], Hanaa Ebrahim [1] and Amr Mokhtar [2]
[1] Switching Department, National Telecommunication Institute, Cairo, Egypt.
[2] Dept. of Electrical Engineering, Alexandria University, Alexandria 21544, Egypt.
{rami.amer@nti.sci.eg, aasherif@alexu.edu.eg, dr.hanaa_nti@yahoo.com, amromokhtar61@gmail.com}

*Abstract*— In this research, we proposes a new method for cooperation and underlay mode selection in cognitive radio networks. We characterize the maximum achievable throughput of our proposed method of hybrid spectrum sharing. Hybrid spectrum sharing is assumed where the Secondary User (SU) can access the Primary User (PU) channel in two modes, underlay mode or cooperative mode with admission control. In addition to access the channel in the overlay mode, secondary user is allowed to occupy the channel currently occupied by the primary user but with small transmission power. Adding the underlay access modes attains more opportunities to the secondary user to transmit data. It is proposed that the secondary user can only exploits the underlay access when the channel of the primary user direct link is good or predicted to be in non-outage state. Therefore, the secondary user could switch between underlay spectrum sharing and cooperation with the primary user. Hybrid access is regulated through monitoring the state of the primary link. By observing the simulation results, the proposed model attains noticeable improvement in the system performance in terms of maximum secondary user throughput than the conventional cooperation and non-cooperation schemes.

## I. INTRODUCTION

Secondary utilization of a licensed spectrum band can enhance the spectrum occupancy and introduce a reliable solution to its scarcity. Secondary users can access the spectrum without introducing harassment to the primary users' performance [1]. In order to achieve cognitive radio objectives, SUs are required to adaptively modify its transmission parameters according to changes in the system or in the PU behavior. Recently, cooperation between the SU and PU has gained a lot of attention in cognitive radio research area. Specifically, SUs act as relays for the PU data while also trying to transmit their own data. In [2], the advantages of the cognitive transmitter acting as a "transparent relay" for the PU transmission are investigated. Authors proved that the stability region of the system increases in terms of the maximum allowed arrival rates of both the PU and SU. Keeping the cooperation as a basis, extensions are made by providing admission control at the SU relay queue. Each timeslot, the SU decide whether or not to accept the PU packet and bear responsibility for PU's packets delivery. Queues' stability and delay in cooperative cognitive radio network with admission control is explained in [3]. A secondary user is equipped with two queues, its own queue and relaying queue, with a battery for energy storage. Primary user's packet is admitted to the relay queue with certain admission probability. Results reveal that, the maximum achievable PU arrival rate is non-decreasing function of the admission probability in the non-energy-constrained system. Additionally, delay encountered by the PU packets is showed to be decreasing function of PU arrival rate with higher delay found in the energy-constrained system.

A network consists of a single cognitive radio transmitter–receiver pair shares .the spectrum with two primary users is proposed in [4]. Each PU has single data queue, whereas the SU has three queues; one storing its own data while the other two are the relay queues corresponding to the relayed packets from PUs. A cooperative cognitive MAC protocol for the proposed network is suggested, where the SU exploits the idle periods of the two PU bands. Traffic arrival at each relay queue is controlled via a tunable admittance factor, while relay queues service is controlled via channel access probabilities assigned to each queue based on the band of operation. The stability region of the proposed protocol is plotted focusing on its maximum expected throughput. The performance gains of the cooperative cognitive protocol with admission control are showed to outperform the noncooperative and conventional cooperative systems. Several contributions imposed the effect of admission control in the study of cooperative cognitive radio networks [5-8].

Sequential decision making in cognitive radio networks has been studied in many works. In [9], the problem of opportunistic spectrum access in time-varying channels with Rayleigh fading is studied, where the channels are characterized by both channel quality and the probability of being occupied by the primary users. A two-dimensional POMDP framework is suggested for opportunistic spectrum access. A greedy strategy is proposed, where a SU selects a channel that has the best-expected data transmission rate to maximize the immediate reward in the current slot. When compare with the optimal strategy that considers current and future rewards, the greedy strategy brings low complexity in design and almost ideal performance. Many research works studied in different perspectives POMDP based sequential decision making in cognitive Radio networks [10-13].

In this paper, we propose a new method of spectrum sharing between primary and secondary users. The SU accesses the channel in a hybrid way; either partially cooperates with the PU or in the underlay mode. This proposed method depends on primary user channel quality and the

**action-reward** model of the system to regulate the admission of each primary packet at the relay queue. We characterize the channel model between each source and destination. We resort to the Finite State Markov Chain (FSMC) model to calculate the transition probabilities between channel SNR levels. The maximum throughput of the SU is plotted for three system models, non-cooperative, cooperative, and the proposed hybrid model.

The rest of this paper is organized as follows. Section II introduces the system model and channel models. Belief and reward functions are found in section III. Section IV presents the performance analysis comparing the different systems under investigation. Finally, concluding remarks are drawn in Section V.

## II. SYSTEM MODEL

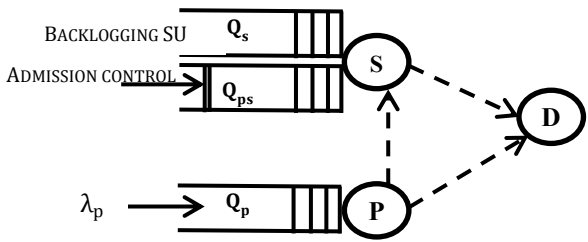

Fig. 1. System model.

Fig. 1 depicts the system model under consideration. The system is composed of one PU and SU transceivers. It is supposed that in a time-varying environment there is one PU transceiver accessing a licensed fading channel with bandwidth B Hz. The SU tries to exploit the channel according to the sensing the channel at each timeslot. Primary user has a data queue, $Q_p$. The arrival process at $Q_p$ is modeled as Bernoulli arrival process with mean $\lambda_p$ [packets/slot]. Considering the SU, it is represented by two data queues, $Q_s$ and $Q_{ps}$. $Q_s$ is an infinite capacity buffer for storing the SU's own packets which is assumed to be backlogged.

It is assumed that the channel quality status of the link between the PU source and destination is calculated from the received SNR at the primary destination. Then the PU destination sends feedback with the received SNR level back to the PU source to obtain the ACK/NACK status corresponding to correct/non-correct transmission, respectively. The PU source obtains the positive ACK from the feedback message if the SNR received at the PU destination was above certain threshold, this will be discussed later. As the SU can overhear the ACK/NACK messages from the PU destination to the PU source, it can send independent ACK feedback message if the PU source-destination link is in outage and the PU packet is correctly received and admitted at the SU source. Additionally, the SU exploits the SNR feedback associated with the ACK/NACK message to build an action-reward model to access the PU channel in the hybrid mode.

We propose a hybrid spectrum sharing works as follows: each timeslot the PU channel is sensed and the status of PU is known. If the PU is sensed to be idle, the SU accesses the channel to serve a packet from $Q_{ps}$ if nonempty otherwise it serves its own data queue. If the channel is sensed as busy, the SU has two choices corresponding to two operation modes. In the underlay mode, and depending on the quality belief function about the primary channel, the SU send its data concurrently with the PU's transmission with small power.
In the cooperative mode, the SU, and based on the belief function, choses to accept the PU packet in the relay queue to bear the responsibility for delivering this packet in case of transmission failure through the direct link. If the PU packet is not received correctly through the direct path and the SU could decode it, the SU will send an ACK message to the PU source to declaring the responsibility for sending the failed packet in subsequent timeslots.

The assumed model of the PU channel is tenable to build the belief function and use it to predict the future of the channel quality, as discussed in next subsections. Furthermore, the SU has a reward function of its decision; this reward is defined as the profit gained by the SU corresponding to the chosen decision. The analysis in the next subsections will show how to map the built belief function to get the appropriate action that maximizes the reward function. Analysis in the next subsections will show how to map the built belief function to get the appropriate action that maximizes the reward function.

*Channel Models*

In this subsection, we explain the channel model of the link between the PU source and destination. All channel models in the network are assumed to have the same model. A fading channel in a time-varying environment is modelled as finite state Markov chain model Fig. 2.

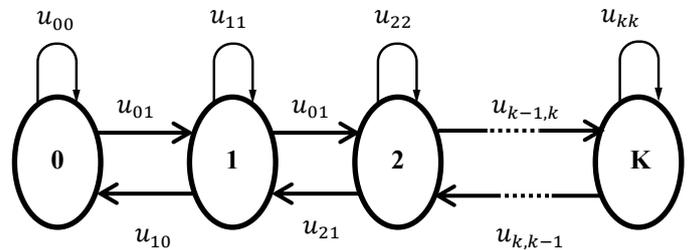

Fig. 2. FSMC channel model

This model is introduced to characterize the channel quality levels of the fading channel, these thresholds are given as:

$$\Gamma_k = e^{k\eta/B} - 1, \qquad k \in \{0, 1, \cdots, K - 1\}, \quad (1)$$

where $0 = \Gamma_0 < \Gamma_1 < \cdots \Gamma_{k-2} < \Gamma_{k-1} = \infty$ and $\eta$ represents the different between adjacent SNR levels in terms of achievable data rate, in units of bps. The source transmits a training sequence in each packet, which is predefined at the receiver, and in turn, the destination receives them and sends feedback to the transmitter with the received SNR [14]. Accordingly, the received instantaneous SNR $\gamma$ is partitioned into $K$ non-overlapping levels. If $\Gamma_k < \gamma < \Gamma_{k+1}$, the current SNR state of the channel is regarded as $k$. The packet transmission is assumed to last for a whole timeslot, and the length of the timeslot is large enough to make the assumption valid, which are the SNR states changes according to the FSMC from certain state only to the adjacent states or regain the same level. In the next paragraphs, we are going to derive an expression for steady state distribution of each level of the quantized channel levels and the rate of crossing each level. These parameters will be used to calculate state transition probabilities between channel quality levels. We will adopt state transition probabilities to build and update the belief function at the SU.

In such a multipath propagation environment with fading channel, the received SNR is a random variable obeys the exponential distribution with the probability density function: $p(\gamma) = \frac{1}{\gamma_0} e^{-\gamma/\gamma_0}$ and **steady state distribution** given by $\pi_k$ of the level $k \in \{0, 1, \cdots, K - 1\}$.

$$\begin{aligned} \pi_k &= \int_{\Gamma_k}^{\Gamma_{k+1}} p(\gamma) \, d\gamma \\ \pi_k &= e^{-\Gamma_k/\gamma_0} - e^{-\Gamma_{k+1}/\gamma_0}, \\ & k \in \{0, 1, \cdots, K - 1\}, \end{aligned} \quad (2)$$

and the **level crossing rate** function is $\Lambda(k)$ for a given level $k \in \{0, 1, \cdots, K - 1\}$:

$$\Lambda(k) = \sqrt{\frac{2\pi \Gamma_k}{\gamma_0}} f_{dopp} e^{-\Gamma_k/\gamma_0}, \quad (3)$$
$$k \in \{0, 1, \cdots, K - 1\}$$

The level crossing rate function $\Lambda(k)$ denotes the average number of times per time interval that the fading signal crosses a certain signal level [15], and $f_{dopp}$ represents the Doppler frequency of the mobile terminal. It is involved here since it was found that the envelope of the fading levels of such a channel is shaped by the Doppler frequency. The adjacent transfer method proposed in [16] is assumed for the transitions between SNR levels, which assumed any transitions only can happen between adjacent states or between the same states:

$$P_{k,l} = 0, \text{ if } |k - l| > 1, \quad (k, l) \in \{0, 1, \cdots, K - 1\}. \quad (4)$$

The state transition probabilities can be obtained directly from the steady state distribution of each level $k \in \{0, 1, \cdots, K - 1\}$ and the level crossing rate function $\Lambda(k)$ of the corresponding level, Therefore, the state transition probabilities can be obtained as follows:

$$\begin{cases} u_{k,k+1} = \frac{\Lambda(\Gamma_{k+1})}{\pi_k} \tau_{pkt}, & k = 0, 1, \cdots, K - 2 \\ u_{k,k-1} = \frac{\Lambda(\Gamma_k)}{\pi_k} \tau_{pkt}, & k = 0, 1, \cdots, K - 1 \end{cases} \quad (5)$$

where $\tau_{pkt}$ represents the packet duration time [14]. Because the summation of each transition probabilities in all states is one, the probability to transit to the same state is given by $u_{k,k}$:

$$u_{k,k} = \begin{cases} 1 - u_{01} & k = 0; \\ 1 - u_{k,k+1} - u_{k,k-1} & k = 2, \cdots, K - 2 \\ 1 - u_{K-1,K-2} & k = K - 1 \end{cases} \quad (6)$$

The decision at timeslots when the PU is active is whether to accept the PU packet in the relay queue, or to access the channel concurrently in the underlay mode. This decision (or action) is built upon the belief function at the SU. The SU built and update a belief function about the probability that the SNR level of the PU channel will be in certain SNR level among $K$ predefined levels. In the next subsection, we introduce the belief function expression and the suggested reward function used to map the probabilities of the belief function to the chosen action. The decision is made to maximize the current reward of a finite horizon problem.

III. BELIEF AND REWARD FUNCTIONS

The SU maintain and update a belief vector representing sufficient statistical information about the PU channel, as the actual quality state is unknown. The SU can know the exact state of the PU channel only when the PU is active as the channel state could be overheard from the feedback message to the PU source. Therefore, an important concept called quality vector is defined to describe the channel quality information and is defined as follows:

$$\Theta_n(t) = [\beta_n^0(t) \ \beta_n^1(t) \ \beta_n^2(t) \ \ldots\ldots \ \beta_n^k(t) \ \beta_n^{K-1}(t)] \quad (7)$$
$$n \in \{0, 1, 2, \ldots, N\}$$
$$k \in \{0, 1, 2, \ldots, K - 1\}$$

where $\Theta_n(t)$ is the belief function about the PU channel quality state in time slot n, and $\beta_n^k(t)$ is the probability that channel quality state in time index n lies in state $k$. It is clear that $\sum_k \beta_n^k(t) = 1$ and the probability $\beta_n^k(t) < 1$.

The update of the belief function each timeslot is dependent on whether the PU is active or not. If the PU is active, the SU listens to the feedback message and obtain the state of the channel quality in that timeslot. If the PU is idle, the update of the belief function is by averaging the belief over all transition probabilities.

$$\beta_n^l(t+1) = \begin{cases} u_{k,l} & \sigma(t) = \text{feedback}, \ \text{level} = k \\ \sum_k \beta_n^k(t) u_{k,l} & \sigma(t) = \text{no feedback} \end{cases} \quad (8)$$

Considering the direct transmission between primary source and primary destination, the achievable rate in this case could be formulated as:

$$\acute{\eta}_P^{DT} = \log_2\{1 + \text{SNR}\} \quad (9)$$

where SNR is the signal to noise ration of the direct transmission between primary source and primary destination, which is assumed before as exponential random variable. The mean of the SNR random variable is $\gamma_{op}$. The target transmission rate of the PU link is $R_p$, which is constant and determined according to the required QoS. The link of direct transmission is said to be non-outage if the target transmission rate of the PU is greater than the achievable rate.

$$P_{DT,\text{no outage}}^p = \Pr\{R_p > \acute{\eta}_P^{DT}\} \quad (10)$$

$$P_{DT,\text{no outage}}^p = \exp\left(-\frac{\rho_p}{\gamma_{op}}\right) \quad (11)$$

where $\rho_p = 2^{R_p} - 1$ and $P_{DT,\text{outage}}^p = 1 - \exp\left(-\frac{\rho_p}{\gamma_{op}}\right)$

For the link of direct transmission to be in non-outage state, the instantaneous SNR value at the desired destination should be greater than certain SNR threshold obtained from (9),

$$\Gamma_{threshold} = 2^{R_p} - 1 \quad (12)$$

The reward function is expressed as an increasing function when the SU tends to cooperate with the PU while the link of direct transmission is in outage. Therefore, the SU is going to admit the PU packets in the relay queue when the direct transmission cannot sustain the required target rate due to the low instantaneous SNR. Additionally, the function is expressed as increasing function when the SU tends to choose the underlay access while the link of direct transmission is not in outage. For every state $\epsilon K$, we have two reward values corresponding to two actions made by the SU; whether to cooperate with PU or access in the underlay mode. We define the reward function as follows.

$$R(k, a_{cooperate}) = \begin{cases} A_k, & \text{SNR} < \Gamma_{threshold} \\ 0, & \text{SNR} > \Gamma_{threshold} \end{cases} \quad (13)$$

$$k \in \{0, 1, 2, \ldots, K-1\}$$

$$R(k, a_{underlay}) = \begin{cases} 0 & \text{SNR} < \Gamma_{threshold} \\ B_k & \text{SNR} > \Gamma_{threshold} \end{cases} \quad (14)$$

$$k \in \{0, 1, 2, \ldots, K-1\}$$

Since the actual reward function cannot be obtained at the beginning of each timeslot due to the randomness of SNR, and the actual state of the channel quality is not known, we refer to the expected reward function $\rho(\Theta_n(t), a)$. The expected reward function is calculated by averaging the reward values over the probabilities that SNR lies in certain state giving the reward value.

$$\rho(\Theta_n(t), a) = \sum_k \beta_n^k(t) \ R(k, a) \quad (15)$$

The expected reward function is calculated for the two possible actions and the action with the highest reward is chosen.

IV. NUMERICAL RESULTS

In this section, a typical scenario is considered in which the channel is slowly fading and the environment is time varying. The performance is evaluated by MATLAB simulations, whose parameters are as follows: Channel bandwidth B = 6 MHz, average received SNR for the PU, $\gamma_{op}$ = 15 dB, carrier frequency fc = 50 MHz, terminal mobile speed $v$ = 2 m/s (Doppler frequency is $\frac{vf}{c}$), data transmission and acknowledgment time $\tau_{da}$ = 100 ms, and the rate interval $\eta$ = 3 Mbps. The system model is established with eight SNR quality levels (K = 8). The average received SNR for the SU, $\gamma_{os}$ = 45 dB. The average received SNR for the SU from PU, $\gamma_{ops}$ = 20 dB. It is assumed that the power of the underlay transmission is decreased by 80 %. The target rate for primary and secondary links, $R_p$, $R_s$, respectively, is 3.5 bps/Hz. As we assume consistent transmission rates over all links. Accordingly, $\Gamma_{threshold}$ $2^{3.5} - 1 = 10.4320$ from (12), this is found to be between the fourth and fifth SNR levels. Number of iteration is 200000.

In Fig. 3, we present the SU throughput versus the primary user arrival rate for the non-cooperation, conventional cooperation, and the proposed hybrid cooperation methods. The figure reveals that the SU throughput of the proposed method is higher than the SU throughput of the non-cooperation and conventional cooperation methods. On the first hand, at low $\lambda_p$, the SU throughput of the proposed hybrid cooperation outperforms the other two methods. This result is owing to the SU exploits the belief function about the PU channel to access the channel in the underlay mode, when the belief function predicts good channel conditions. On the other hand, when $\lambda_p$ becomes higher, the SU throughput of the proposed method is only the non-zero value as the SU could access the channel only in the underlay mode concurrently with the PU

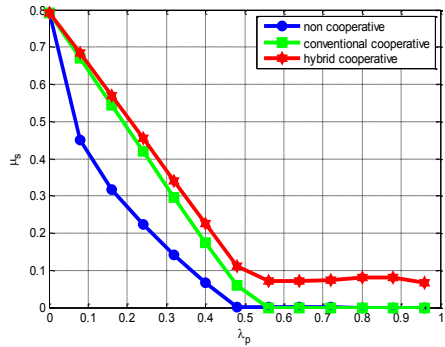

Fig. 3. System SU throughput versus the PU arrival rate.

Figure. 4 shows the maximum SU throughput for different underlay transmission powers. Secondary user throughput is plotted versus three different power levels, 0.2, 0.4, and 0.6 Watt. The SU transmission power in the overlay mode is 1 Watt. Result reveals that increasing the underlay power of SU leads to lower SU throughput when $\lambda_p$ is small and higher SU throughput when $\lambda_p$ is high. This can be interpreted as, at low $\lambda_p$, the SU with higher underlay power causes more interference to the PU. The more interference at the PU results in more retransmission of PU undelivered packets either through the direct link or by the relaying SU. This leads to lower opportunities for the SU to access the channel in the overlay mode to serve his data. On contrary, the SU mainly accesses the channel in the underlay mode when $\lambda_p$ is high, so the higher underlay power gives the better SU throughput

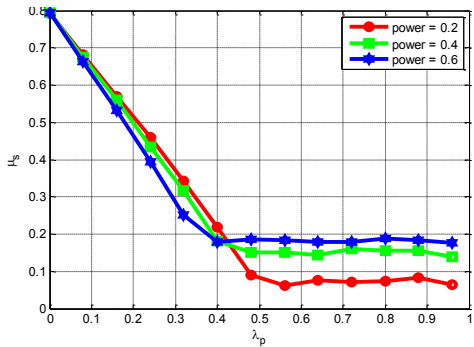

Fig. 4. SU throughput for different SU underlay power

## V. Conclusion

This paper illustrates cooperation and underlay mode selection in a cognitive radio network. The SU either cooperates with the PU to relay its data, or access the channel concurrently with the PU in the underlay mode. We assumed a cooperative system with admission control at the SU relay queue. Mode selection operation is based on the current state of the PU channel to its destination. Specifically, the admission is function of the channel quality of the direct link between primary source and destination. The system model is shown and the mathematical analysis is presented using the FSMC model. The performance analysis and simulation shows the improvement in network performance.